\documentclass[a4,useAMS,usenatbib]{mn2e}

\usepackage[english]{babel} 
\usepackage{subfigure}
\usepackage{graphicx}

\usepackage[fleqn]{amsmath} 
\usepackage{color}

\usepackage[varg]{txfonts}
\usepackage{url}

\newcommand{\noi}{\noindent}
\newcommand{\vect}{\bmath}

\newcommand{\msun}{M_\odot}
\newcommand{\lsun}{L_\odot}
\newcommand{\kms}{\, {\rm km\, s}^{-1}}

\newcommand{\Mpc}{\, {\rm Mpc}}
\newcommand{\lsystem}{SDSS~J120602.09+514229.5}

\title[\lsystem]{Quantifying dwarf satellites through gravitational imaging: the case of \lsystem}

\author[S. Vegetti, O. Czoske \& L.V.E. Koopmans]{%
 Simona Vegetti\thanks{E-mail: vegetti@astro.rug.nl}, 
  Oliver Czoske\thanks{Current address: Institut f\"ur Astronomie,
T\"urkenschanzstr.~17, A-1180~Wien, Austria} \& L\'eon V. E. Koopmans\\
  Kapteyn Astronomical Institute, University of Groningen, PO Box 800, 9700\,AV
  Groningen, the Netherlands}
  
\begin{document}

\date{Accepted  ... Received ...; in original form ....}

\pagerange{\pageref{firstpage}--\pageref{lastpage}} \pubyear{2002}

\maketitle

\label{firstpage}

\begin{abstract}
\lsystem~ is a gravitational lens system formed by a group of galaxies at redshift $z_{\rm{FG}}=0.422$ lensing a bright background galaxy at redshift $z_{\rm{BG}}=2.001$. The main peculiarity of this system is the presence of a luminous satellite near the Einstein radius, that slightly deforms the giant arc. This makes \lsystem ~the ideal system to test our grid-based Bayesian lens modelling method, designed to detect galactic satellites independently from their mass-to-light ratio, and to measure the mass of this dwarf galaxy despite its high redshift.
We model the main lensing potential with a composite analytical density profile consisting of a single power-law for the group dominant galaxy, and two singular isothermal spheres for the other two group members.
Thanks to the pixelized source and potential reconstruction technique of \citet{Vegetti09a} we are able to detect the luminous satellite as a local positive surface density correction to the overall smooth potential.
Assuming a truncated Pseudo-Jaffe density profile, the satellite has a mass $M_{\rm{sub}}=(2.75\pm0.04)\times10^{10}\msun$ inside its tidal radius of $r_t=0.68\arcsec$. This result is robust against changes in the lens model, with a fractional change in the substructure mass from one model to the other of 0.1 percent. We determine for the satellite a luminosity of $L_{B} = (1.6\pm0.8)\times10^{9}\,\lsun$, leading to a total mass-to-light ratio within the tidal radius of $(M/L)_{B} = (17.2\pm8.5)\msun / \lsun$. The central galaxy has a sub-isothermal density profile as in general is expected for group members. From the SDSS spectrum we derive for the central galaxy a velocity dispersion of
$\sigma_{\mathrm{kinem}} = 380\pm60\,\mathrm{km\,s^{-1}}$ within the
SDSS aperture of diameter $3\arcsec$. The logarithmic density slope of $\gamma=1.7^{+0.25}_{-0.30}$ (68\% CL), derived from this measurement, is consistent within 1-$\sigma$ with the density slope of the dominant lens galaxy $\gamma\approx 1.6$ determined from the lens model. This paper shows how powerful pixelized lensing techniques are in detecting and constraining the properties of dwarf satellites at high redshift.
\end{abstract}


\section{Introduction}
Comparison between numerical CDM simulations and direct
observations of the Milky Way and Andromeda has shown the
existence of a strong discrepancy in which the abundance of predicted subhaloes
outnumbers that of observed dwarf galaxies \citep[e.g.][and references therein]{Kauffmann93,Klypin99,Springel08,Kravtsov09}. Reconciling the
luminosity function with the mass function is therefore a crucial test for CDM models. 
With the specific aim of addressing this issue, a grid-based Bayesian lens modelling code was developed by \citet{Vegetti09a}.  This technique is able to identify possible mass substructure in the lensing potential, 
by reconstructing the surface brightness distribution of lensed arcs and Einstein rings.
Several tests on mock data have shown that we can detect mass substructure as massive as $M_{\rm{sub}}\ge 10^8\msun$ \citep{Vegetti09a,Vegetti09b}.\\
In a recent application to the lens system SDSSJ0946+1006~from the Sloan Lens ACS Survey (SLACS, \citet{Bolton06}), the method has proved to be successful in recovering 
the smooth lensing potential and in identifying a satellite with a high mass-to-light ratio, ${\rm (M/L)}_{{\rm V}}\ga 120~ (\rm {M/L})_{\rm{V},\odot}$, and
with mass $M_{\rm{sub}}\sim(3.51\pm0.15)\times 10^9\msun$, while reconstructing the data to the noise level \citep{Vegetti10}.    
However, the complexity of the data and systematic effects related for example to sub-pixel structure, PSF modelling and spatially varying noise 
can complicate the source and the potential reconstruction and all their effects always have to be carefully assessed and quantified. A definitive test is therefore required to \emph{calibrate} the capability of our technique. 
\lsystem~\citep[][]{Lin09} with a luminous satellite right on the lensed images (see fig. 1) is an ideal system to accomplish this task. 
In this paper we present a full analysis of  \lsystem, measuring the mass and the mass-to-light-ratio of the dwarf satellite and showing the strength of the method on known satellites.
The layout of the paper is as follows. In Section 2 we introduce the data. In Section 3 we provide a short description of the modelling method and a detailed description of the main results and in Section 4 we summarise our main results.  Throughout the paper we assume the following cosmology $H_0 = 73\,\kms\Mpc^{-1}$, $\Omega_{\rm m}=0.25$ and $\Omega_\Lambda=0.75$.

\section{The data}
\lsystem~ (the Clone) was observed with the Hubble Space Telescope (HST) in
cycle~16 (P.I.: S.~Allam). WFPC2 images were obtained through three
filters, F450W, F606W, and F814W.  We base our lens model on the F606W
data because they provide the best combination of depth and
resolution. \\
Four dithered exposures were obtained, each with an integration time
of 1100\,s. We retrieved the calibrated exposures from the HST archive
and used \textit{multidrizzle} to combine them. The final image has a
pixel scale of $0.05\,\mathrm{arcsec}$.\\
To model the observed structure of the lensed source, we require
knowledge of the point spread function (PSF) of the drizzled
image. Since there are no suitable stars in the field, we rely on a
model PSF created with Tiny Tim, v6.3
\citep{Krist1993}.\footnote{\url{http://www.stsci.edu/software/tinytim/tinytim.html}}
The PSF is generated for the position of the lens system on chip~2 of
the WFPC2, subsampled by a factor 10. Subsampling is necessary because
the dither pattern involves half-pixel shifts which cannot be taken
into account by Tiny Tim alone.  Instead we rebin the highly
subsampled PSF to the original WFPC2 pixel scale once for each science
exposure, with the output grid shifted by five subsampled pixels to
account for half-pixel shifts.  The rebinned PSF is inserted at the
position of the lens galaxy G1 in copies of the four science files
after setting the image data to zero. These PSF exposures are then
drizzled with the same parameters as before to create an approximation
to the PSF of the drizzled science image.\\
Light from the outer parts of the main lens galaxies G1, G2 and G3
contributes a few percent to the light at the location of
the lensed arc images. We subtract de Vaucouleur models of these
galaxies, generated with \textsc{galfit}\footnote{\url{http://users.obs.carnegiescience.edu/peng/work/galfit/galfit.html}}, version 3.0 \citep{Peng2002}. The effective radius of G1 is determined at 3.9$\pm$0.1 arcsec.\\
Finally, we need to remove the satellite galaxy G4. Since this galaxy
sits on top of an arc image and it is not \textit{a priori} clear what
the background level due to light from the arc is, no unambiguous
model of G4 can be determined. Due to its compactness we opt to
subtract a simple Gaussian model with FWHM $0.173\,\mathrm{arcsec}$ and
normalisation determined from visual impression of the image after
subtraction. Changes in the normalisation within a plausible range do
not significantly affect the potential reconstruction.\\
A spectrum of G1 is available from the Sloan Digital Sky Survey. This
spectrum has a fairly low signal-to-noise ratio of $S/N\approx 8$ (per
pixel) and no kinematic measurements are given in the SDSS database.
With the template fitting method described in Czoske et al.~(2010, in
preparation; see also \citealt{Czoske2008}) and the Indo-US spectrum
\citep{Valdes2004} of the K2III star HD~195506 as template we obtain a
good fit (reduced $\chi^2 = 1.11$; Fig. \ref{fig:spectrum}) with a velocity dispersion of
$\sigma_{\mathrm{kinem}} = 380\pm60\,\mathrm{km\,s^{-1}}$ within the
SDSS aperture of diameter $3\arcsec$. This is lower but not
inconsistent with the value that \citet{Lin09} obtained from fitting
a singular isothermal ellipsoid model to the lens configuration,
$\sigma_{\mathrm{lens}} = 440\pm7\,\mathrm{km\,s^{-1}}$. Note,
however, that this value applies to the entire mass doing the lensing,
whereas our value is for the main lens galaxy G1 only.

\begin{figure}
   \begin{center} 
      \includegraphics[width=7cm]{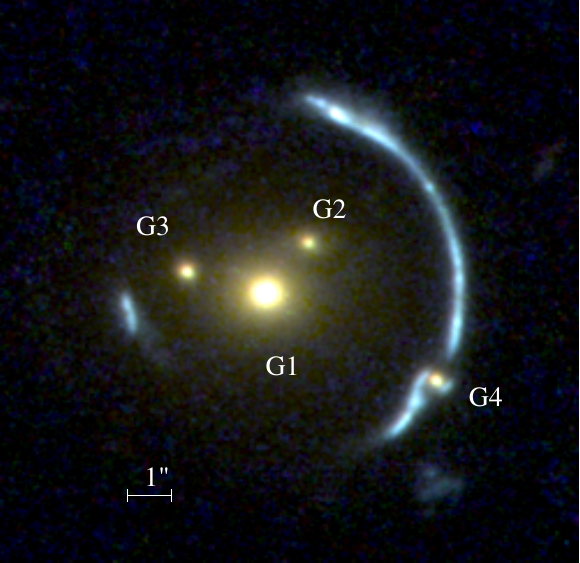}
      \caption{Overview of the lens system \lsystem. This false-colour image was created from HST/WFPC2 images in F450W, F606W and F814W.}
      \label{fig:overview} 
    \end{center}     
 \end{figure}	

\begin{figure*}
   \begin{center} 
      \includegraphics[width=\hsize]{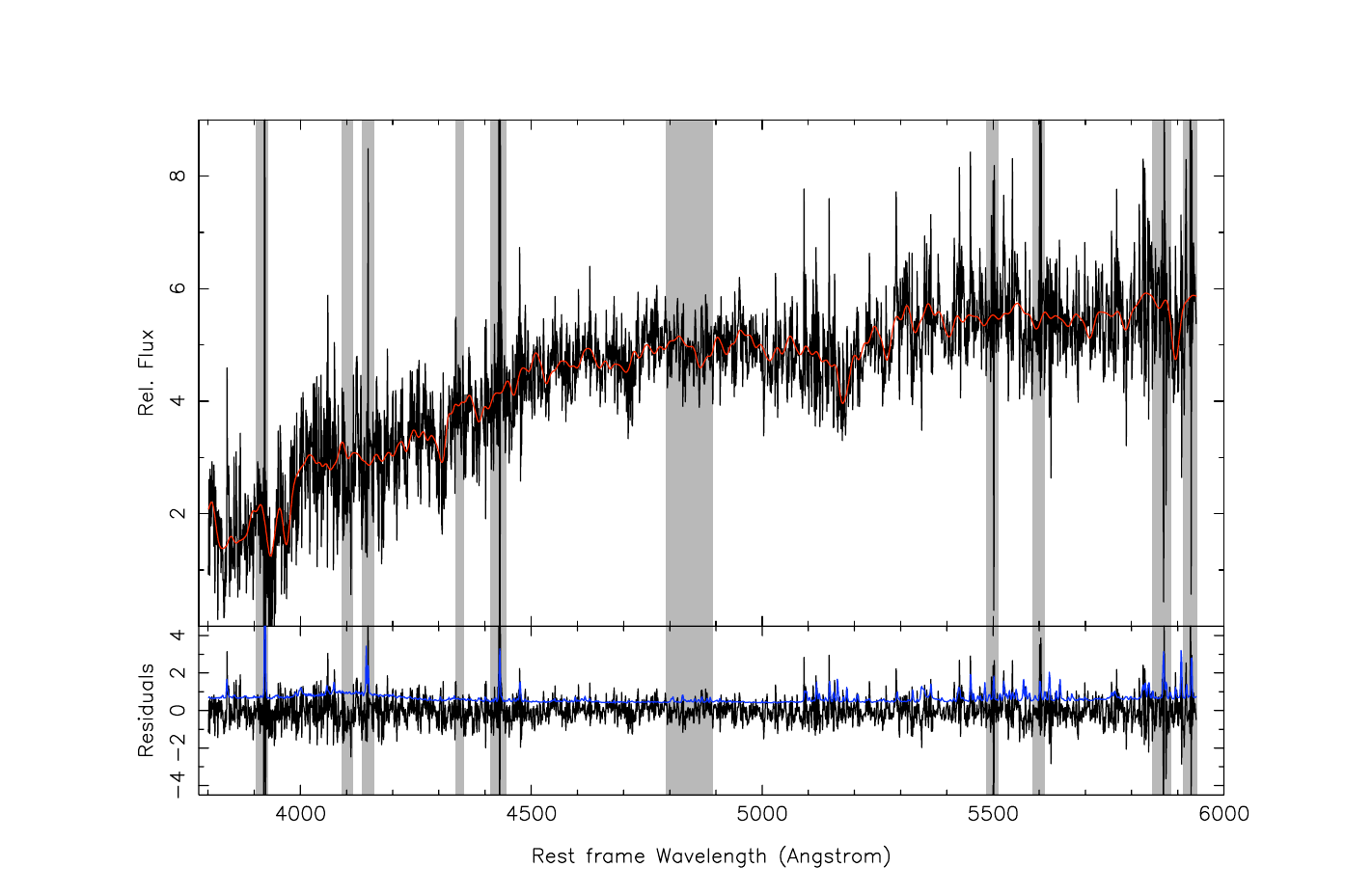}
      \caption{SDSS spectrum of galaxy G1 is shown in the upper panel. The spectrum of 
K2III star HD~195506 is convolved with a Gaussian line-of-sight velocity 
distribution of dispersion $380\,\mathrm{km\,s^{-1}}$ overlaid in red. The 
residuals of the fit are plotted in the lower panel, with the expected 
noise spectrum overlaid in blue. The masked regions are for Balmer lines, 
metal lines that typically show abundance anomalies compared to galactic 
template spectra, atmospheric absorption features and regions that appear 
contaminated with strong spikes.}
      \label{fig:spectrum} 
    \end{center}     
 \end{figure*}	
 
\section{Lens Modelling}
The lens modelling is performed using the Bayesian adaptive method of \citet{Vegetti09a} to which we refer for a detailed description. Briefly we proceed as follows:

\begin{enumerate}
\item  Initially, we only assume a smooth analytic model for the main lens potential (i.e. we only consider G1, G2 and G3) and maximize the relative posterior probability in terms of their lens parameters. At this point we ignore the satellite G4.
\item We fix the lens potential at the maximum posterior values found in the previous iteration and run a simultaneous pixelized reconstruction of the source surface brightness distribution $\vect{s}$ and the potential correction $\delta\vect{\psi}$. This leads to the detection and localisation of possibly present mass substructures in the lens potential.
\item Finally, we build a composite analytic model in which both the main lenses and the detected satellite have a   power-law (PL) density profile and we optimize the relative penalty function for the corresponding parameters. 
The power-law is truncated in the case of the satellite.
\end{enumerate}
In the next sections we describe this procedure in more detail as applied to the lens \lsystem, and show that it is able to detect and quantify the satellite G4.

\subsection{Smooth potential parametric reconstruction}
We initially start with a smooth model that explicitly excludes the satellite G4. We model the lensing potential as the combination of a single power-law ellipsoid for G1 and two singular isothermal spheres (SIS) for G2 and G3 with a surface density in terms of critical density $\Sigma_c$ \citep{Kormann94}
\begin{equation}
\Sigma\left(r\right) = \frac{\Sigma_c~b}{2\sqrt q~r^{\gamma-1}}\,,
\end{equation}

\noi where $r=\sqrt{x^{2} + y^{2}/q^{2}}$.\\
Given the relatively high dynamic range of the lensed image surface brightness distribution, the source is reconstructed on a Delaunay tesselation grid that is built from the image plane by casting every second pixel in RA and DEC back to the source plane. The area of each triangle (i.e. the grid resolution) depends on the local lens magnification. A curvature source regularisation is adopted \citep{Vegetti09a}. The free parameters for the posterior probability maximization are the lens strength $b$, the position angle $\theta$, the axis ratio $q$ and the density slope $\gamma$ for G1, the lens strength for G2 and G3, the strength of the external shear $\Gamma_{\rm{sh}}$ and its position angle $\theta_{\rm{sh}}$.
The best PL+2SIS model is reported in Table 1. We find that this is an incomplete and simplified description for the true lensing potential of \lsystem~and do not expect it to provide a good description of the data. It does however provide a sufficient starting point for the next modelling step. Following \citet{Lin09} we also tried a simplified model where the lensing potential of G1, G2 and G3 is described by a single global SIE (plus external shear), with free parameters being $b$, $\theta$, $q$, the centroid coordinates $x_c$ and $y_c$, $\Gamma_{\rm{sh}}$ and $\theta_{\rm{sh}}$ (see Table 1). For this model our results are consistent with \citet{Lin09} but still provide an approximate description of the lens data. Assuming the total mass inside the Einstein radius and the SDSS luminosity-weighted stellar velocity dispersion as two independent constraints, we determine an effective logarithmic density slope, based on spherical Jeans modelling (see Koopmans et al. 2006 for details), of $\gamma=1.7^{+0.25}_{-0.30}$ (68\% CL), assuming orbital isotropy ($\beta=0$), an effective radius of 3.9$\pm$0.1 arcsec for G1 and a seeing of 1.5 arcsec. Conversely, the best PL density slope of G1, i.e.\ $\gamma=1.58$, predicts a dispersion of 340 km/s. Hence the best PL model is in excellent agreement with the measured stellar velocity dispersion, although the SIE model, which predicts a large dispersion of 450 km/s is still marginally in agreement given the larger error on its measured value.

\begin{table*}
\caption{ Best recovered parameters for the mass model distribution for the lens \lsystem. For each of the considered models we report the best recovered set of non-linear parameters and the galaxy centroids.}
\label{tab:results}
\begin{center}
\begin{tabular}{ccccccccccc}
\hline Model&Lens& $b$/$M_{\rm{sub}}$& $\theta$&$q$&$x_{\rm{c}}$&$y_{\rm{c}}$&$\gamma$&$\Gamma_{\rm{sh}}$&$\theta_{\rm{sh}}$&$\log\cal{L}$\\ 
&&&(deg)&&(arcsec)&(arcsec)&&&(deg) \\ 
\hline 
PL+2SIS&G1&1.59&-50.7&0.75&$-$0.54&$-$0.13&1.48&$-$0.06&$-$27.8&48956.90\\
&G2&0.54&&$\equiv1.00$&0.38&$-$0.98&$\equiv2.0$\\
&G3&0.55&&$\equiv1.00$&$-$2.23&0.33&$\equiv2.0$\\
\\
SIE&G1+G2+G3&3.67&$-$71.2& 0.76&$-$0.45&0.002&$\equiv2.0$&0.005&$-$63.1&50267.43\\
\\
PL+2SIS+PJ&G1&2.27&$-$79.2&0.80&$-$0.54&$-$0.13&1.58&0.03&3.53&102308.08\\
&G2&0.17&&$\equiv1.00$&0.38&-0.98&$\equiv2.0$\\
&G3&0.13&&$\equiv1.00$&$-$2.23&0.33&$\equiv2.0$\\
&G4&$2.78\times10^{10}\msun$&&&3.12&$-$2.10&\\
\\
SIE+PJ&G1+G2+G3&3.78&-74.7&0.80&$-$0.45&0.002&$\equiv2.0$&0.02&$-$74.1&114035.73\\
&G4&$2.75\times10^{10}\msun$&&&3.12&$-$2.10\\
\hline
\end{tabular} 
\end{center}
\end{table*}

\subsection{Satellite detection}\label{sec:non_parametric}
\begin{figure*}
   \begin{center} 
      \includegraphics[width=\hsize]{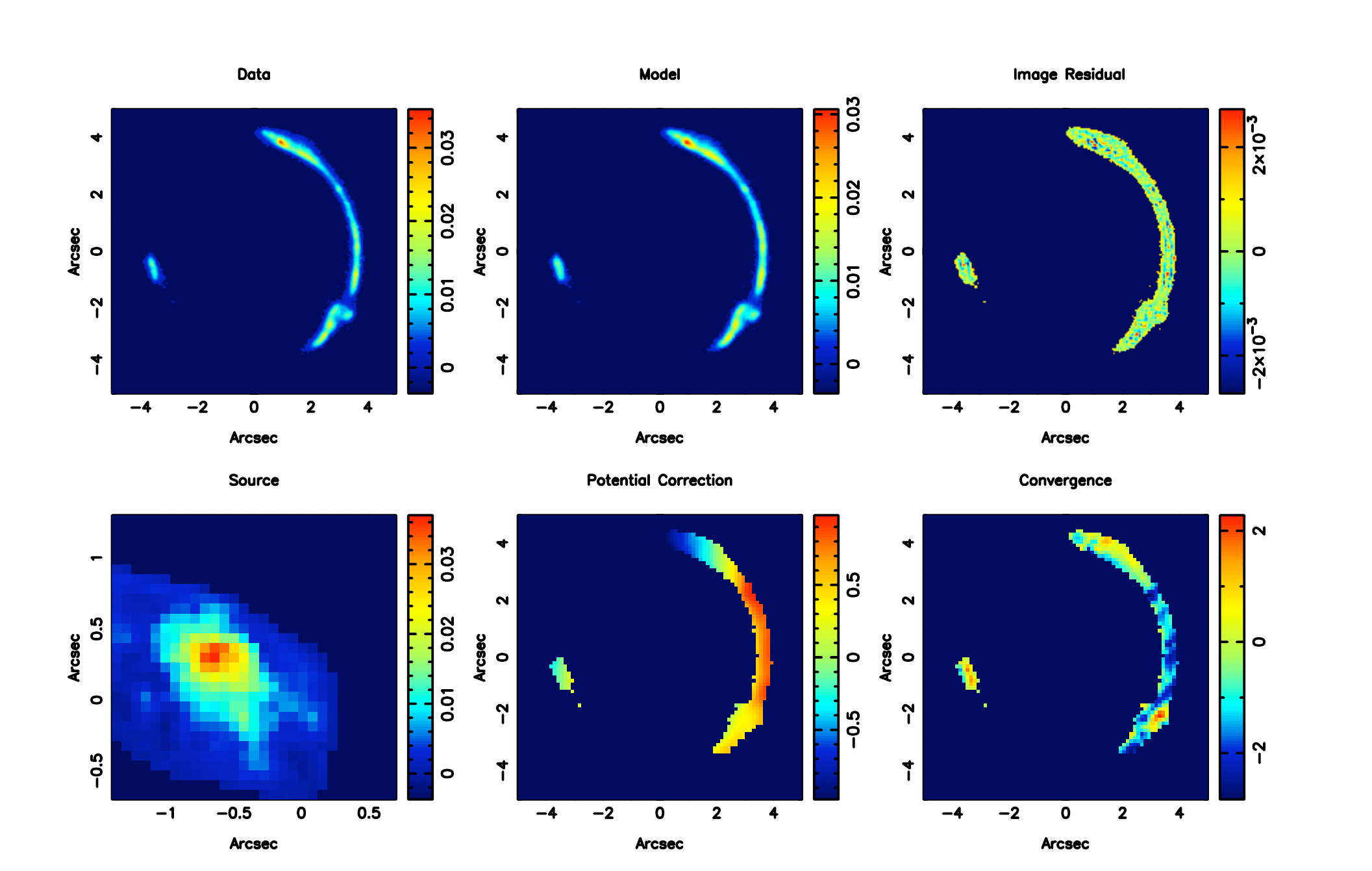}
  \caption{Results of the pixelized reconstruction of the source and lens potential corrections. The top-left panel shows the original lens data, the middle one shows final reconstruction while the top-right one shows the image residuals. On the second row the source reconstruction (left), the potential correction (middle) and the potential correction convergence (right) are shown.}
      \label{fig:potcorr} 
    \end{center}     
 \end{figure*}

\begin{figure*}
   \begin{center} 
      \includegraphics[width=1\hsize]{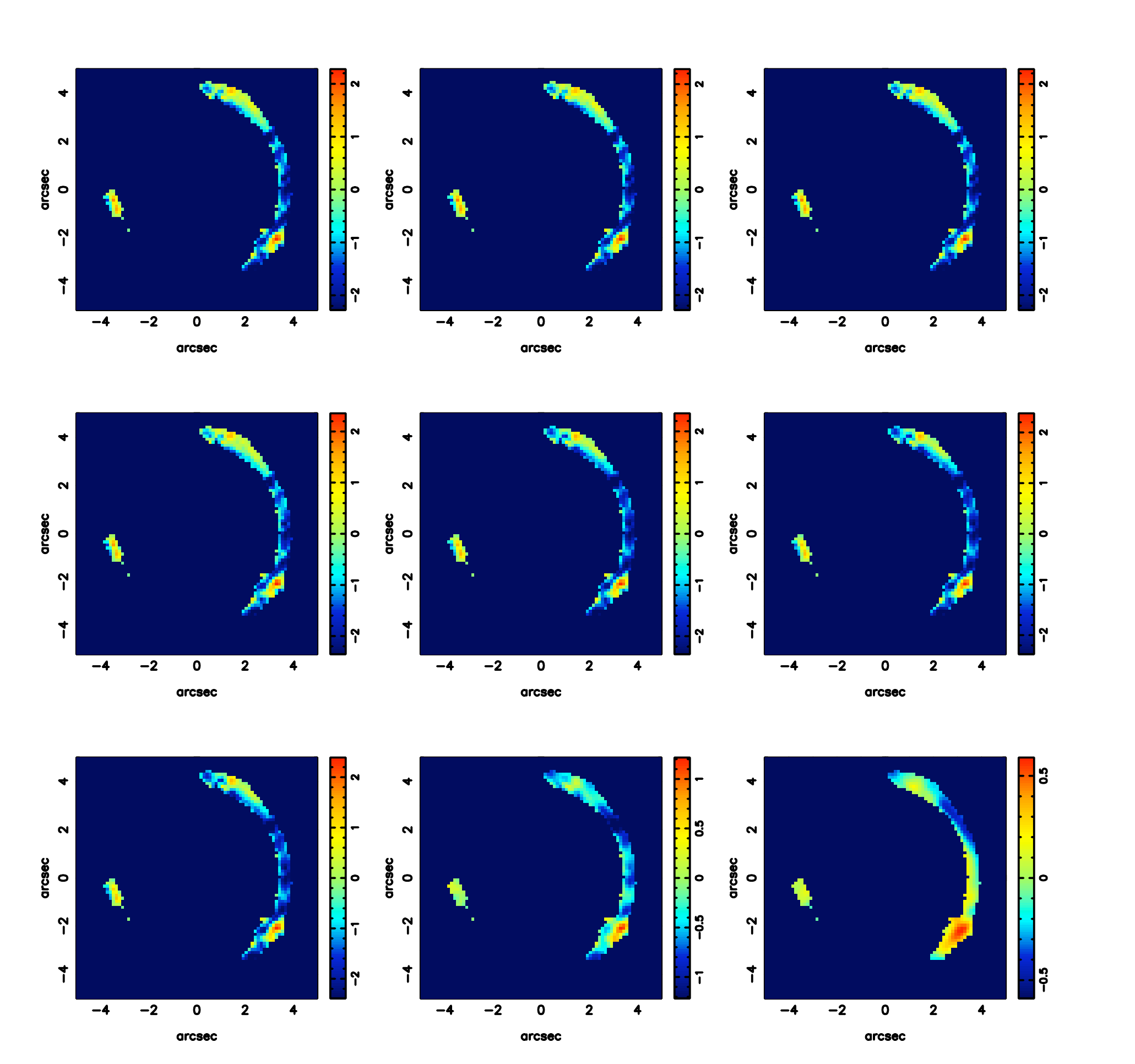}
\caption{Results of the pixelized reconstruction of the convergence corrections for different values of the  source and  potential regularization $\lambda_{s}=3\times10^3$ (top row), $\lambda_{s}=3\times10^4$ (middle row), $\lambda_{s}=3\times10^6$ (bottom row) $\lambda_{\delta\psi}=3\times10^7$ (left column) and $\lambda_{\delta\psi}=3\times10^8$ (middle column) and $\lambda_{\delta\psi}=3\times10^9$ (right column).}
      \label{fig:test} 
    \end{center}     
 \end{figure*}

The next step is to test whether the pixelized technique is able to identify the satellite G4.
We fix the lens parameters of G1, G2 and G3 to the values found in the previous section and run a linearized grid-based reconstruction for the source surface brightness and lens potential corrections. To ensure the linearity of the solution, both the source and the potential corrections are initially over-regularized and then the relative regularization constants are slowly lowered. Curvature regularization is used for both the source $\vect{s}$ and potential corrections $\delta\vect{\psi}$.\\
The potential corrections are reconstructed on a regular Cartesian grid with $81\times81$ pixels and a pixel scale of $0.12\arcsec$. Via the Poisson equation $\delta\vect{\psi}$ can be translated into convergence (surface density) corrections $\delta\vect{\kappa}=\frac{1}{2}(\delta\psi_{11}+\delta\psi_{22})$. A strong positive convergence correction is found at the exact position of G4 (see Fig. \ref{fig:potcorr}). 
Smooth non-negligible density corrections are also found on the upper side of the arc. These could be related to the fact that the source regularization is not at its optimal level, but slightly over-regularized, or more likely to smooth deviations of the starting model from the true mass distribution \citep[see e.g.][]{Barnabe09}. \\
Reconstructions with different values of source and potential regularization lead to very similar results (see fig. \ref{fig:test}). As expected, the potential correction at the position of G4 becomes more extended and less concentrated for higher levels of regularizations ($\lambda_s=3.0\times10^6$ and $\lambda_{\delta\psi}=3.0\times10^9$, $\lambda_s=3.0\times10^6$ and $\lambda_{\delta\psi}=3.0\times10^8$) but is otherwise the same for all other combinations of these parameters. This indicates the robustness of the results against changes in the source structure and potential smoothness.\\
Also the single global SIE leads to a similar convergence map, with the density correction corresponding to G4 located at the same position and having a comparable intensity as for the multiple-component model. 
The satellite detection is therefore robust against different choices for the initial smooth global lens potential. In fact a SIE+PJ model is slightly better than a PL+2SIS+PJ one. This could be interpreted as due to the presence of a common halo for this group of galaxies. It is important to note that the convergence correction is located exactly at the position of the peak of the surface brightness distribution of G4 as recovered in Section 2 via a Gaussian fit.\\
We conclude that the extra freedom allowed to the lens potential via the linear potential corrections compensates/corrects for the inadequacies of the global lens potential and both identifies and precisely locates possible mass substructure.

\subsection{Satellite mass}
 In this section we further quantify the pixelized substructure by an analytic model and constrain the relative parameters in the context of that model.
We assume an analytic mass model consisting of a single PL for G1, two SIS for G2, G3 and a Pseudo-Jaffe (PJ) for G4 as well as a simplified model SIE+PJ. 
The Pseudo-Jaffe profile reads as \citep{Dalal02,Vegetti10}:
\begin{equation}
\Sigma(r) = \frac{\Sigma_c~b_{\rm{sub}}}{2}\left[r^{-1}-(r^2+r_t^2)^{-1/2}\right]\,,
\end{equation}
where $r_t =\sqrt{b_{\rm{sub}}b}$ is the tidal radius for a lens strength $b_{\rm{sub}}$.
The satellite G4 is centred on the position where the peak of the convergence correction was found by the pixel-based reconstruction. The free parameters for G1, G2, and G3 are the same as before, while the only free parameter for G4 is the mass within the tidal radius $M_{\rm{sub}}=\pi r_t b_{\rm{sub}} \Sigma_c$. The recovered best parameters are listed in Table 1 for both models. The inferred substructure mass is not strongly affected by small changes in the substructure position; a systematic change of 1 pixel in the centre coordinates leads, for example, to a change in the substructure mass of only 1 percent.\\
They respectively lead to a satellite mass and tidal radius $M_{\rm{sub}}=(2.78\pm0.04)\times10^{10}\msun$, $r_t=0.68\arcsec$ (PL+2SIS+PJ) and $M_{\rm{sub}}=(2.75\pm0.04)\times10^{10}\msun$, $r_t=0.81\arcsec$ (SIE+PJ).\\
The reader should not be tempted to compare the different models in terms of the Likelihood reported in Table 1; models can only be compared in terms of the Bayesian evidence, which requires to integrate over the multidimensional  space of the posterior probability density distribution over the free-parameters. The model comparison is not relevant for our current analysis and this step is  therefore not carried out.\\

\subsection{Satellite mass-to-light ratio}
Finally, we estimate the luminosity of G4 by integrating the Gaussian model to the F606W
surface brightness profile obtained in Sect.~2. We expect this to lead to a
underestimate of the luminosity, because of the sharply dropping wings of the Gaussian model.
Fitting more realistic models is, however, difficult due to the compactness of G4 and the contamination with the arc light. 
The colour of G4 is consistent with that of the main lens galaxies G1, G2 and G3, which
indicates an old stellar population. The absolute rest-frame B-band
magnitude is obtained following the prescription of \citet{Treu1999}
for an elliptical galaxy and is $M_{B} = -17.5$, corresponding to a
luminosity of $L_{B} = (1.6\pm0.8)\times10^{9}\,\lsun$. The large error estimate
includes the uncertainty due to arc light contaminating G4 and the model
uncertainty for the light profile. The total mass-to-light ratio of G4, inside the tidal radius, is thus $(M/L)_{B} = (17.2\pm8.5)\msun / \lsun$.
As explained, this should be really only considered an upper limit to the true mass-to-light ratio. Plausible de Vaucouleur profiles
are typically 0.8 mag brighter than the Gaussian model, leading to a total luminosity of  $L_{B} = (3.3\pm1.6)\times10^{9}\,\lsun$ and a
mass-to-light ratio of $(M/L)_{B} = (8.2\pm2.6)\msun / \lsun$. This result
is consistent with little to no dark matter inside the tidal radius of this satellite; this is also in agreement with the typical stellar mass-to-light ratio at this redshift $(M/L)_{B}\approx5\msun / \lsun$ \citep{Treu05}.
 
\section{Summary}
We have applied the grid-based Bayesian lensing code by \citet{Vegetti09a} to the lens system \lsystem, which has a known luminous satellite located on the lensed arc. We have shown that the perturbation of the lensed arc, created by the satellite, can be used to gravitationally identify the satellite itself and determine its lensing properties, in particular to get an accurate mass measurement. We performed several tests that show that the satellite detection and its recovered mass are robust against changes in the source structure, level of lens potential smoothness and choice of the smooth global lensing model. The main results of this work can be summarised as follows:
\begin{itemize}
\item  A relatively complex model, containing one single power-law, two singular isothermal spheres and a Pseudo-Jaffe satellite, yields a satellite mass $M_{\rm{sub}}=(2.75\pm0.04)\times10^{10}\msun$ inside the tidal radius. This result is consistent with a simpler SIE+PJ model.
\item  The satellite has a total mass-to-light ratio within the tidal radius of $(M/L)_{B}\approx 8.0\msun/\lsun$,
consistent with the presence of little to no dark matter inside the tidal radius, assuming a typical stellar $(M/L_B)_{\star}\approx 5.0\msun/\lsun$
\item  G1, the main galaxy in the group, has a density profile which is sub-isothermal with slope $\gamma=1.58\pm0.1$. This is not unexpected for galaxies in groups \citep[e.g][]{Sand04}
\item  We measure for G1 a velocity dispersion of $\sigma_{\mathrm{kinem}} = 380\pm60\,\mathrm{km\,s^{-1}}$ within the SDSS aperture of diameter $3\arcsec$. This is consistent with the $\sigma_{\rm SIE}$ value from \citet{Lin09} obtained by fitting a singular isothermal ellipsoid model to the lens configuration. From a more proper lensing and dynamics model we predict a stellar velocity dispersion of 340 km/s for the best PL model of G1 that as a logarithmic density slope of $\gamma=1.58$. Conversely, we predict a density slope of $\gamma=1.7^{+0.25}_{-0.30}$ (68\% CL) from the observed stellar velocity dispersion. This agrees very well with that determined from the PL model of G1, but is also still marginally in agreement with the SIE model.
\end{itemize}

This paper demonstrates the great potential of pixelized lensing techniques in robustly identifying and measuring the key properties of small mass structure/dwarf satellites in distant galaxies.
The application of this method to a large uniform set of lens galaxies will allow in the near future to constrain the general properties of mass substructure in galaxies and to test the CDM paradigm on these small scales.

\section*{Acknowledgements}
SV, OC and LVEK are supported (in part) through an NWO-VIDI program subsidy (project number 639.042.505)
Based on observations made with the NASA/ESA Hubble Space Telescope, obtained at the Space Telescope Science Institute, which is operated by the Association of Universities for Research in Astronomy, Inc., under NASA contract NAS 5-26555.
\bibliography{ms}
\clearpage

\end{document}